# Enhanced Harmonic Generation in Terahertz FELs: Influence of Pre-Bunching and Undulator Geometry on Spectral and Angular Emission

A. A. MOLAVI CHOOBINI[1], S. S. GHAFFARI-OSKOOEI[2], F. FARAHI[3] AND F. M. AGHAMIR[1, *]

[1]*Dept. of Physics, University of Tehran, Tehran 14399-55961, Iran.*
[2]*Department of Atomic and Molecular Physics, Faculty of Physics, Alzahra University, Tehran, Iran.*
[3]*Department of Physics, university of Ottawa, Ottawa, Ontario, Canada.*

*aghamir@ut.ac.ir

**Abstract:** Both theoretical and numerical analyses are conducted to investigate terahertz (THz) radiation emission from free-electron lasers. The angular and spectral characteristics of radiation are analytically evaluated leveraging Liénard–Wiechert field formalism, The analysis spanned across varying beam profiles and undulator parameters, including harmonic order, magnetic strength, period, and electron energy. Fourier analysis of the electric field reveals strong sensitivity of spectral content to underlying beam dynamics and undulator structure. Planar undulators tend to generate radiation with more prominent side lobes, especially at higher harmonics, while helical undulators produce narrower, more focused emission profiles. Importantly, the polarization and forward-directed nature of the radiation are governed primarily by the resonant conditions and beam dynamics, rather than the undulator geometry itself. Building on these analytical insights, GENESIS simulations are employed to quantify the evolution of the bunching factor and radiation intensity under varying energy spread and plasma dispersion. The results show that increased undulator periods enhance radiation coherence, while transitions to wiggler regimes (at high magnetic parameters) lead to angular broadening and lateral lobe formation. Among all beam profiles studied—Gaussian, Lorentzian, bi-Gaussian, and pre-bunched—the pre-bunched configuration demonstrates superior harmonic generation capabilities, albeit with high sensitivity to energy spread and plasma effects.

## 1. Introduction

The terahertz (THz) frequency range has emerged as a vibrant area of research, driven by its unique ability to penetrate optically opaque materials while remaining non-ionizing—making it ideal for non-destructive imaging and probing [1–5]. THz waves also exhibit high sensitivity to molecular vibrations, rotational transitions, and collective excitations, enabling powerful spectroscopic applications across physics, chemistry, biology, and materials science [6, 7]. Realizing coherent, high-power THz sources, however, remain a central challenge. Among the candidate technologies, free electron lasers (FELs), particularly in their self-amplified spontaneous emission (SASE) configuration, have demonstrated exceptional potential. FELs offer tunable, high-intensity, and coherent radiation across a wide spectral range, including the THz domain, enabling ultrafast studies of matter at unprecedented spatial and temporal resolution [8, 9]. Their operation is based on the microbunching of relativistic electron beams within magnetic structures known as undulators, which convert spontaneous emission into coherent radiation. The geometry of these undulators—planar or helical—critically shapes the angular, spectral, and polarization characteristics of the emitted THz radiation [10, 11]. As such, FELs provide a powerful and versatile platform for generating and tailoring THz waves for both fundamental research and applied technologies.

Significant advances have been made in developing THz radiation sources based on free-electron lasers (FELs), with diverse approaches targeting enhanced coherence, efficiency, and spectral control. Weihao Liu et al. proposed and investigated a scheme for generation of near-terahertz frequencies via Smith–Purcell radiation from dual gratings in a free-electron laser (FEL) configuration [12]. Suresh C. Sharma and colleagues explored the interaction of laser-modulated relativistic electron beams with surface plasma waves, yielding THz radiation through nonlinear current densities acting as antenna-like sources [13]. Guo-Qian Liao et al. produced high-intensity THz pulses by irradiating metal foils with picosecond lasers, showcasing the ability to manipulate the THz spectrum by adjusting laser pulse duration or target size [14]. Klaus Floettmann et al. explored Cherenkov–wakefield radiation as a potential THz source in FELs, emphasizing electron beam interaction with dielectric-lined narrow tubes [15].

Advancements in THz FEL diagnostics have also emerged: M. Lenz and co-workers investigated measuring broadband terahertz FEL radiation pulses using electro-optic sampling [16]. Their findings suggest that highly efficient beam-to-radiation energy conversion can be achieved by employing a strongly tapered helical undulator operating at the zero-slippage resonant condition. Eléonore Roussel et al. introduced a dual-output electro-optic sampling system for high-resolution signal retrieval at the European X-FEL, bridging photonic time-stretch and phase-diversity techniques [17]. A. Fisher's work with circular waveguides and tapered undulators demonstrated efficient beam–radiation synchronization and energy extraction with approximately 10% energy efficiency [18], while Vittoria Petrillo proposed synchronized THz and soft X-ray emission from FEL oscillators using energy recovery linacs [19]. Other studies, such as that by K. Zhukovsky and colleagues analyzed the influence of electron beam parameters on radiation in FELs, comparing their results with experimental data in the THz region [20]. Their study highlights how emittance, energy spread, and beam current density together affect the radiation characteristics of a single-pass FEL. Leon Feigin and his team investigated a high-power terahertz-free electron laser using tapering-enhanced super radiance [21]. Their unique undulator design incorporates a tapered undulator under the zero-slippage condition, resulting in a notably more potent and efficient THz radiation source. Andrew Fisher et al. demonstrated simultaneous dual-frequency lasing in compact THz FELs using pre-bunched beams [22].

These foundational efforts underscore the central importance of beam dynamics, undulator design, and coherent emission mechanisms in THz FEL systems—core themes pursued in this study. Building on this foundation, a unified analytical–numerical framework for characterizing and optimizing THz emission from FELs is presented. Employing an exact Liénard–Wiechert field formulation and generalized Bessel expansion, angular-spectral distributions, coherence properties, and harmonic generation across a range of beam profiles and undulator configurations is investigated. This formulation is complemented by a novel application of the bunching factor as a phase-coherence diagnostic, which bridges spontaneous emission theory with coherent radiation through microbunching, modelled across a spectrum of beam distributions. The synergy between analytical modeling and simulations reveals that pre-bunched electron beams, despite their susceptibility to energy spread and plasma-induced decoherence, offer unparalleled harmonic generation efficiency. The present work introduces a comparative analysis of the bunching factor across Gaussian and Lorentzian profiles, demonstrating that Lorentzian beams exhibit superior resilience to plasma-induced decoherence, enabling more robust beam preferences for broadband THz applications. This positions them as a transformative platform for narrowband THz emission in attosecond pulse generation, quantum material diagnostics, and ultrafast spectroscopy. Modulation of the undulator field strength enhances the beam–field coupling, thereby increasing the radiation output and overcoming conventional design limitations of compact free-electron laser sources. The study also delineates the transition to the wiggler regime, where angular broadening and the formation of sidelobes become prominent. Distinct advantages of undulator configurations are clarified: helical undulators produce circularly polarized radiation with reduced sidelobes, ideal for applications in chiral molecular spectroscopy and magneto-optics, while planar undulators provide linearly polarized emission with simpler implementation. The bi-Gaussian profile introduces controllable interference-driven modulations, while the Lorentzian profile demonstrates robustness against coherence degradation, enriching the beam design toolkit. In addition, results are benchmarked against longitudinal THz generation via transition-Cherenkov mechanisms in laser-produced plasma filaments, establishing performance boundaries and complementarity between FEL-based and plasma-based THz sources.

The paper is organized as follows: In section II, the theory of THz radiation mechanism is presented. Discussion on THz fields and properties of THz emission are given in section III. Conclusions are drawn in section IV.

## 2. Theoretical Mechanism

There are two common schemes of undulators with variable polarization and mutually orthogonal fields. Figure 1 illustrates a schematic depiction of the interaction between the electron beam and plasma within the crossed helical undulator, with the central region indicating the beam–plasma interaction zone. The trajectories of electrons entering the undulator with imperfect initial conditions will diverge from the desired paths. Thus, fluctuations in the resonance condition will occur concerning a co-propagating light wave. The intricate motion from imperfect injection diminishes the connection to the optical fundamental frequency and induces emission and amplification in higher harmonics. In the framework of the helical undulator, consider a beam of electron bunches propagating in a helical undulator in which the magnetic field is described as:

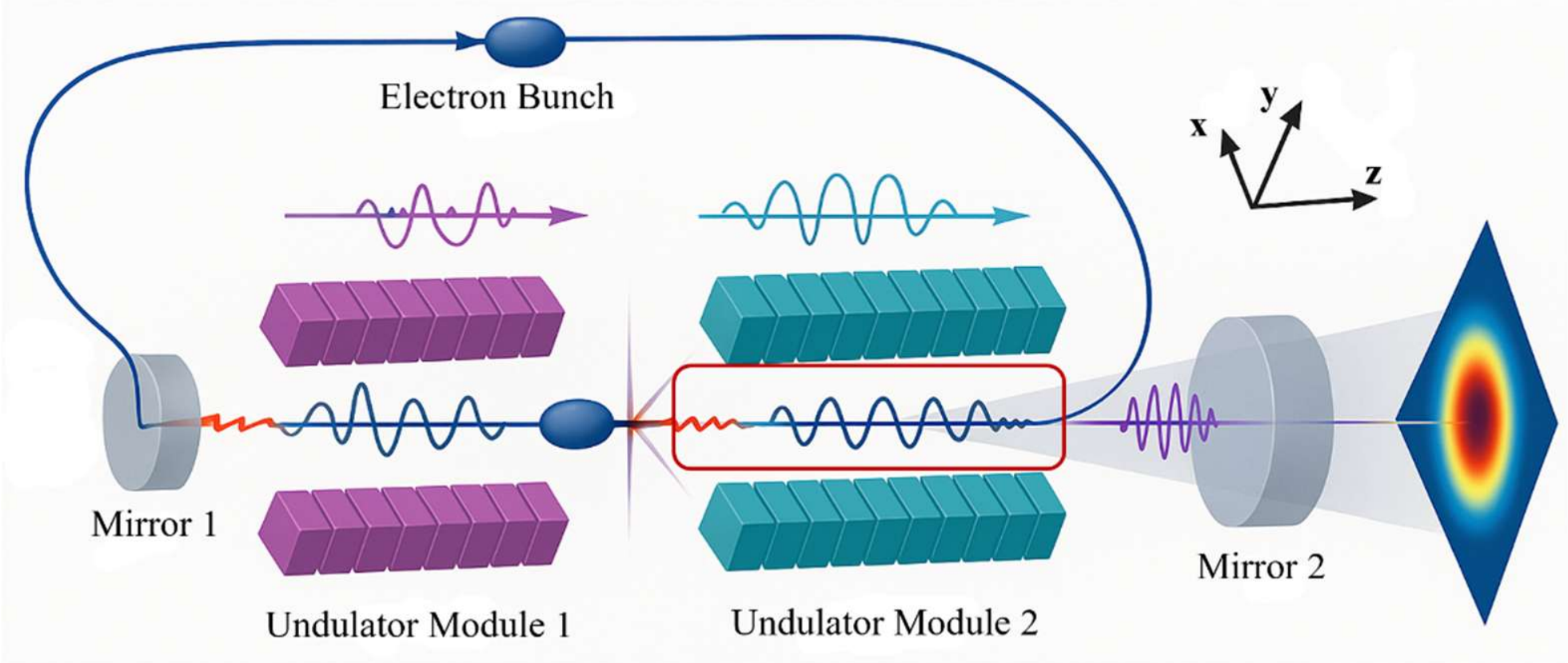

**Fig. 1.** Schematic illustration of laser–plasma interaction in a crossed helical undulator.

$$\vec{B} = B_0[\mathrm{Cos}(k_u z)\,\hat{e}_x + \mathrm{Sin}(k_u z)\,\hat{e}_y] \tag{1}$$

where $B_0$ is the peak field strength and $k_u$ is the wave number of undulator field in transverse plane. In a FEL setup, the electron beam can experience microbunching, where electrons bunch together at certain phases of the radiation field. The emitted radiation spectrum of any electron is determined by its velocity or trajectory as it travels through the undulator. The velocity of electrons can be evaluated by solving the momentum equation and is expressed as:

$$\vec{v} = -\frac{\mathrm{c}a_{\mathrm{u}}}{\gamma}\mathrm{Cos}(k_u z)\,\hat{e}_x - \frac{\mathrm{c}a_{\mathrm{u}}}{\gamma}\mathrm{Sin}(k_u z)\,\hat{e}_y + \sqrt{1 - \frac{a_u^2+1}{\gamma^2}}\,\hat{e}_z \tag{2}$$

where $a_u = \frac{eB_0}{m_e k_u c^2}$ is the normalized magnitude of the magnetic field. From another perspective, this parameter represents the attributes of both gain spectra and spontaneous emission in higher harmonics. Transverse deflections induce radiation in the forward direction. For the investigation of the angular and spectral features of FEL radiation, the general form of total radiation energy per solid angle is given by:

$$\frac{dW}{d\Omega} = \int_{-\infty}^{+\infty} \left|\vec{A}(t)\right|^2 dt \tag{3}$$

here $\vec{A}(t)$ is the vector potential [23]. Through Parseval's theorem, the radiation energy within a specific frequency interval $d\omega$ and solid angle $d\Omega$ can be expressed as:

$$\frac{\partial^2 W}{\partial\omega\partial\Omega} = 2\left|\vec{A}(\omega)\right|^2 \tag{4}$$

The Fourier transform of vector potential $\vec{A}(\omega)$, is attained by Lienard-Wiechert fields:

$$\vec{A}(\omega) = \left(\frac{e^2}{8\pi^2 c}\right)^{\frac{1}{2}} \int_{-\infty}^{\infty} e^{i\omega\left(t - \frac{\hat{e}_r \,.\, \vec{r}(t)}{c}\right)} \frac{\hat{e}_r \times \left(\hat{e}_r - \vec{\beta}\right) \times \dot{\vec{\beta}}}{\left(1 - \vec{\beta}.\hat{e}_r\right)^2} dt \tag{5}$$

Where $\vec{\beta} = \frac{\vec{v}}{c}$ and $\hat{e}_r = Sin\theta Cos\phi \hat{e}_x + Sin\theta Sin\phi \hat{e}_y + Cos\theta \hat{e}_z$ are normalized velocity and observation unit vector, respectively. By insertion of vector potential into Eq. (4), the radiated energy per frequency per solid angle can be verified as:

$$\frac{\partial^2 W}{\partial\omega\partial\Omega} = \left(\frac{e^2\omega^2}{4\pi^2 c}\right)\left|\int_{-\infty}^{\infty} e^{i\omega\left(t - \frac{\hat{e}_r \,.\, \vec{r}(t)}{c}\right)} \hat{e}_r \times \left(\hat{e}_r - \vec{\beta}\right) dt\right|^2 \tag{6}$$

Evaluation of $\frac{\partial^2 W}{\partial\omega\partial}$ in a helical undulator for small angle approximation is contingent upon verification of the integral in the time interval T, where $T = \frac{N_b \lambda_u}{c}$ indicates the undulator transit time. Here $N_b$ and $\lambda_u$ are the undulator total number of periods and wavelength, respectively. Furthermore, the integral must be multiplied by the factor $N_e^2 |b|^2$, where $N_e$ and b are the number of electrons and bunching factor. The microbunching mechanism converts the spontaneous emission of electrons (described by Lienard-Wiechert fields) to a coherent emission. A standard configuration of relativistic electron beams propagating through either helical or planar undulators, as illustrated in Figure 1, is assumed. The FEL interaction is modeled using the Liénard–Wiechert fields for single-electron radiation and is extended to collective effects through the bunching factor |b|, which quantifies the microbunching induced by the self-amplified spontaneous emission (SASE) process. The FEL gain and amplification are incorporated via the evolution of |b| in GENESIS 1.3 simulations, where the three-dimensional FEL equations self-consistently resolve the beam–radiation coupling, including both energy modulation and density bunching along the undulator. This approach goes beyond simple coherent emission from pre-bunched beams by capturing the full, noise-initiated, dynamical interaction. The bunching factor |b|, defined as $|b| = \left|exp\left(ik_r z_j\right)\right|$ over macroparticles j, quantifies phase coherence within $\lambda_r$ and appears in the mathematical expression of the radiated energy. In GENESIS 1.3, it is computed via Fourier analysis of the longitudinal density distribution at each undulator step, incorporating space-charge and radiation fields. Therefore, the following expression can be derived through expansion of the exponential in Eq. (6) and use of generalized Bessel function $J_n(x, y; \tau)$ [24]:

$$\frac{\partial^2 W}{\partial\omega\partial\Omega} = \frac{N_e^2 |b|^2 T^2 e^2}{4\pi^2 c} \sum_{odd-n} A_n \left[\frac{T}{2}\omega Sinc(\omega H - n\omega_u)\right]^2 \tag{7}$$

here $A_n$ is defined as follows:

$$A_n = |Sin\theta \, \Delta_n + F\Delta_{n+1} + F^*\Delta_{n-1}|^2 + |F\Delta_{n+1} + F^*\Delta_{n-1}|^2 \tag{8}$$

where $F = \frac{a_u}{4\gamma} e^{i(\phi + \frac{\pi}{4})}$, $\Delta_n = e^{-in\left(\frac{u}{2} + \frac{\pi}{4}\right)} J_n\left(-\sqrt{u^2 - d^2}, 0; e^{iu}\right)$, $u = tg^{-1}(2tg\phi)$, $\alpha = \frac{a_u^2}{4}\omega \, \mathrm{Sin}\theta$, $d = \frac{a_u Sin\theta Sin\,\phi}{\gamma\omega_u}$, $\omega_u = \frac{2\pi c}{\lambda_u}$, $H = 1 - \xi Cos\theta$, $\xi = 1 - \frac{1 + \frac{a_u^2}{4}}{2\gamma^2}$,

and $J_n(x, y; \tau)$ is the nth order Bessel function of the first kind. The detailed derivation of Eq. (6) is provided in Appendix A.

In the context of linearly polarized undulators, the pathways traversed by electrons exhibit a higher degree of complexity compared to those in helical configurations. Despite precise injection procedures, electrons experience rapid oscillations along the $z$-axis, leading to the phenomenon of spontaneous emission occurring predominantly in odd higher harmonics. To address this challenge, the magnetic field generated by a planar undulator in close proximity to the axis is considered as:

$$\vec{B} = B_0\,\mathrm{Sin}(k_u z)\,\hat{e}_y \tag{9}$$

Under ideal injection conditions, the equations for Lorentz force have exact solutions, yielding the velocity and the trajectory of electrons, respectively:

$$\vec{v} = -\frac{\mathrm{c}a_\mathrm{u}}{\gamma}\,\mathrm{Cos}(k_u z)\,\hat{e}_x \tag{10a}$$

$$\vec{r}(t) = -\frac{a_u\lambda_u}{\sqrt{2}\pi\gamma}Sin(\omega_u t)\hat{e}_x + \left[ct + \frac{a_u^2\lambda_u}{8\pi\gamma^2}Cos(2\omega_u t)\right]\hat{e}_z \tag{10b}$$

Within the planar undulator configuration, the distribution of intensity lacks the azimuthal symmetry observed in the helical configuration. Consequently, for the small radiation angle, the energy emitted per unit solid angle and frequency bandwidth can be expressed as follows:

$$\frac{\partial^2 W}{\partial\omega\partial} = \frac{8N_e^2 e^2\gamma^2}{4\pi^2 c}\sum_{n=1}^{\infty}\left(\frac{2n\xi}{a_u}\frac{Sin\delta_n}{\delta_n}\right)^2\left[\frac{\gamma^2 Sin^2\theta}{2a_u}\Psi_{0,n}^2 + \frac{\sqrt{2}\gamma Sin\theta}{a_u}Cos\varphi\Psi_{0,n}\Psi_{1,n} + \Psi_{1,n}^2\right] \tag{11}$$

where $\Psi_{i,n}$ $(i = 0,1)$ is defined as:

$$\Psi_{i,n} = (-1)^{n+i}\sum_{m=-\infty}^{\infty}(-1)^m J_m(2n\xi)\,[J_{n-2m-i}(n\Xi) + J_{n-2m+i}(n\Xi)] \tag{12}$$

and other parameters in Eq. (11) are, $\delta_n = N_u\pi\left[n - \frac{\omega}{2\gamma^2\omega_u}(1 + a_u^2 + \gamma^2 Sin^2\theta)\right]$, $\Xi = \sqrt{2}\Theta Cos\varphi$, $\Theta = \frac{2a_u\gamma Sin\theta}{1+a_u^2+\gamma^2 Sin^2\theta}$, $\xi = \frac{a_u^2}{1+a_u^2+\gamma^2 Sin^2\theta}$, and $N_u$ is number of periods.

According to Equation (12), the width of individual emission line in the spectrum is dependent upon the number of periods $N_e$ in the argument of $Sin\delta_n/\delta_n$ and THz radiation can occur in a narrow range of wavelengths satisfying $\delta_n \approx 0$. The properties of the emitted THz waves are critically dependent on the magnitude of the parameter $a_u$. For $a_u < 1$, only small harmonic numbers significantly contribute to the THz radiation, while for $a_u > 1$ larger number of harmonics participate. For $a_u \approx 1$ the energy in the fundamental increases and the first few harmonics also possess comparable intensity. When $a_u \gg 1$, numerous closely spaced harmonics are present, and the THz radiation spectrum closely resembles the wide-ranging synchrotron emission.

Building upon the single-electron radiation framework, this study extends the analysis to realistic electron beams with various longitudinal charge distributions. While Eqs. (6)–(12) describe spectral and angular emission from individual electrons, coherent FEL radiation emerges from collective effects such as microbunching and phase synchronization—both strongly influenced by the beam's longitudinal profile and reflected in the bunching factor $|b|$. To capture these multi-particle dynamics, a numerical investigation of four representative beam distributions: Gaussian, Lorentzian, bi-Gaussian, and pre-bunched is conducted. Each profile introduces unique spatial correlations that significantly modify the emission predicted by single-particle theory. Analytical forms of the bunching factor and energy spread suppression are incorporated into radiation integrals, while GENESIS 1.3 simulations reveal the evolution of harmonic content. This combined approach illustrates how beam structure and undulator parameters jointly determine FEL coherence, efficiency, and spectral control, offering critical insights for optimizing compact THz sources under realistic operating conditions.

## 3. Results and Discussion

One of the available sources of THz radiation is free electron laser which can be utilized by reduction of undulator period, proposed in Ref. [25]. The Lienard-Wiechert fields, which can best describe the radiation fields of accelerated charges, are used in the present study for the evaluation of the angular and spectral dependence of radiation. The analysis results in the derivation of the quantity $\partial^2 W_{THz}/\partial\omega\partial\Omega$, which defines the angular and spectral dependency of radiation and is influenced by both undulator and beam parameters.

Understanding the generation of coherent THz radiation in free-electron lasers (FELs) requires a multi-scale approach that bridges the microscopic dynamics of individual electrons with the collective behavior of electron ensembles. While single-electron simulations provide insight into the fundamental mechanisms of radiation emission -- including resonance effects, harmonic generation, and the role of key parameters like undulator strength and beam energy -- practical FEL systems operate with complex, distributed electron beams whose spatial and energy profiles strongly influence coherence and spectral characteristics. The present study examines radiation behavior observed in single-electron models alongside electron ensemble analyses, highlighting how ensemble properties—such as bunching, modulation, and energy spread—shape the efficiency, bandwidth, and harmonic structure of the emitted radiation.

The simulated scheme is FEL in THz regime, with radiation from micro-bunched beams in undulators, including harmonic generation and plasma effects. All calculations assume the following baseline parameters unless varied: electron energy $\gamma = 10 - 50$, undulator period $\lambda_u = 1 - 5\ cm$, magnetic parameter $a_u = 0.5 - 5$, energy spread $\Delta\gamma/\gamma = 0.1 - 1\%$, and number of periods $N_u = 50 - 100$. GENESIS simulations use a 3D grid with $10^5$ macroparticles, and analytical results are derived for small-angle approximation. In single-electron model, numerical simulations of THz radiation patterns for both helical and planar undulators are shown in Figures 2a through 2d, where waveform peaks arise from constructive interference. Increasing the number of undulator periods strengthens electron–photon interactions, producing more intense radiation through resonance effects. Figures 2a and 2b illustrate the transition from undulator to wiggler behavior as the magnetic parameter $a_u$ increases from values below unity to well above it. This transition narrows the THz radiation bandwidth and introduces pronounced lateral lobes, particularly in planar undulators, due to greater transverse electron motion at higher $a_u$. While a larger $a_u$ generally boosts the emitted radiation, nonlinear saturation effects eventually limit the gain, with the impact being especially evident in planar configurations. Figures 3a and 3b show that both helical and planar undulators generate only odd harmonics on axis ($\theta = 0$). Higher-order harmonics increase amplitude but reduce efficiency and introduce sidelobes and angular shifts due to phase mismatch and enhanced transverse motion. The effect of beam energy (represented by the Lorentz factor $\gamma$) is examined in Figures 3c and 3d. As $\gamma$ increases, the radiation cone narrows, amplitude rises, and sidelobes diminish, consistent with relativistic Doppler effects. These findings on the radiation cone agree with earlier results in transition-Cherenkov radiation studies [26].

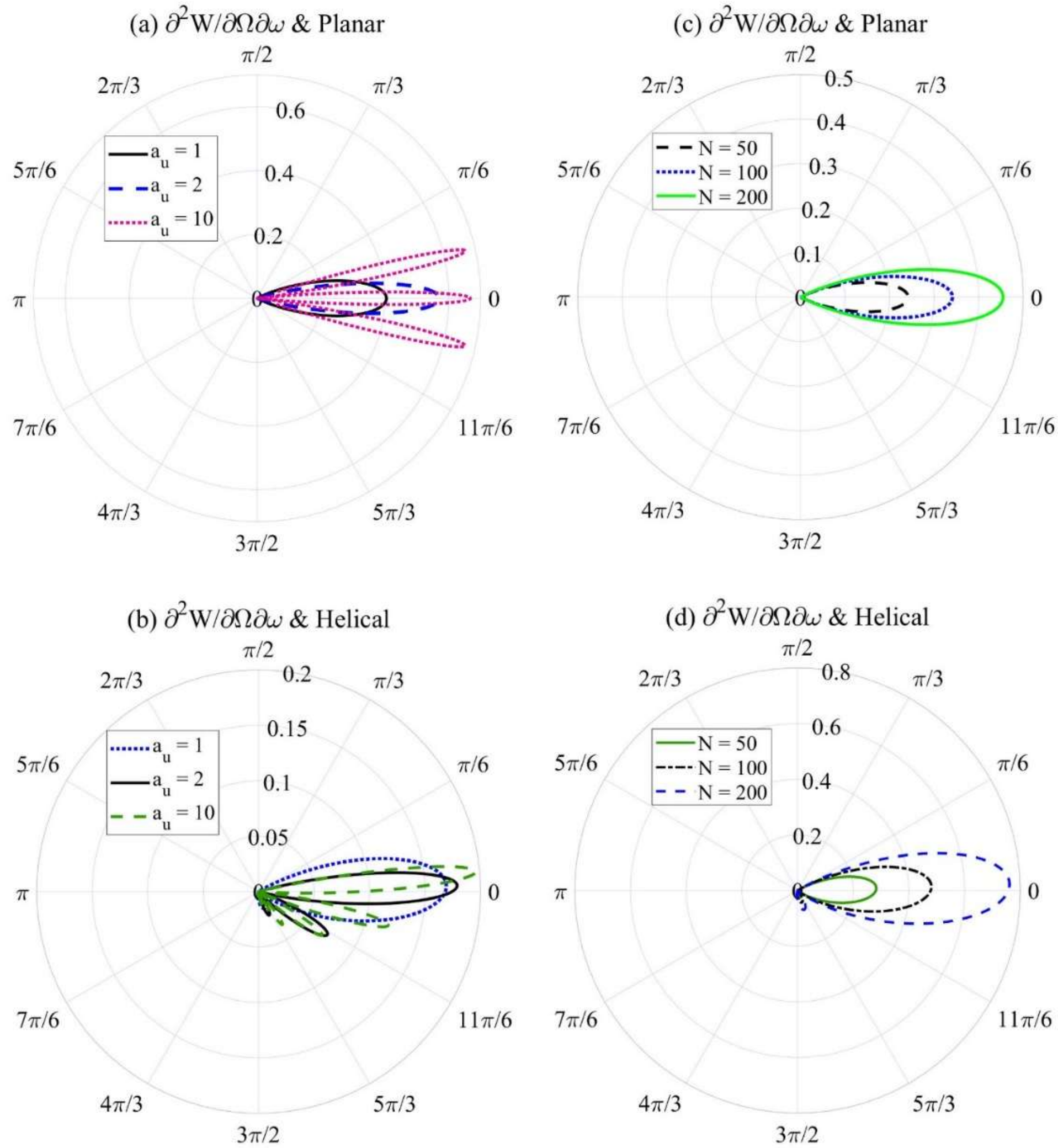

**Fig. 2.** FEL radiation patterns for different values of the magnetic parameter ($a_u$) and varying total numbers of periods in planar and helical undulators.

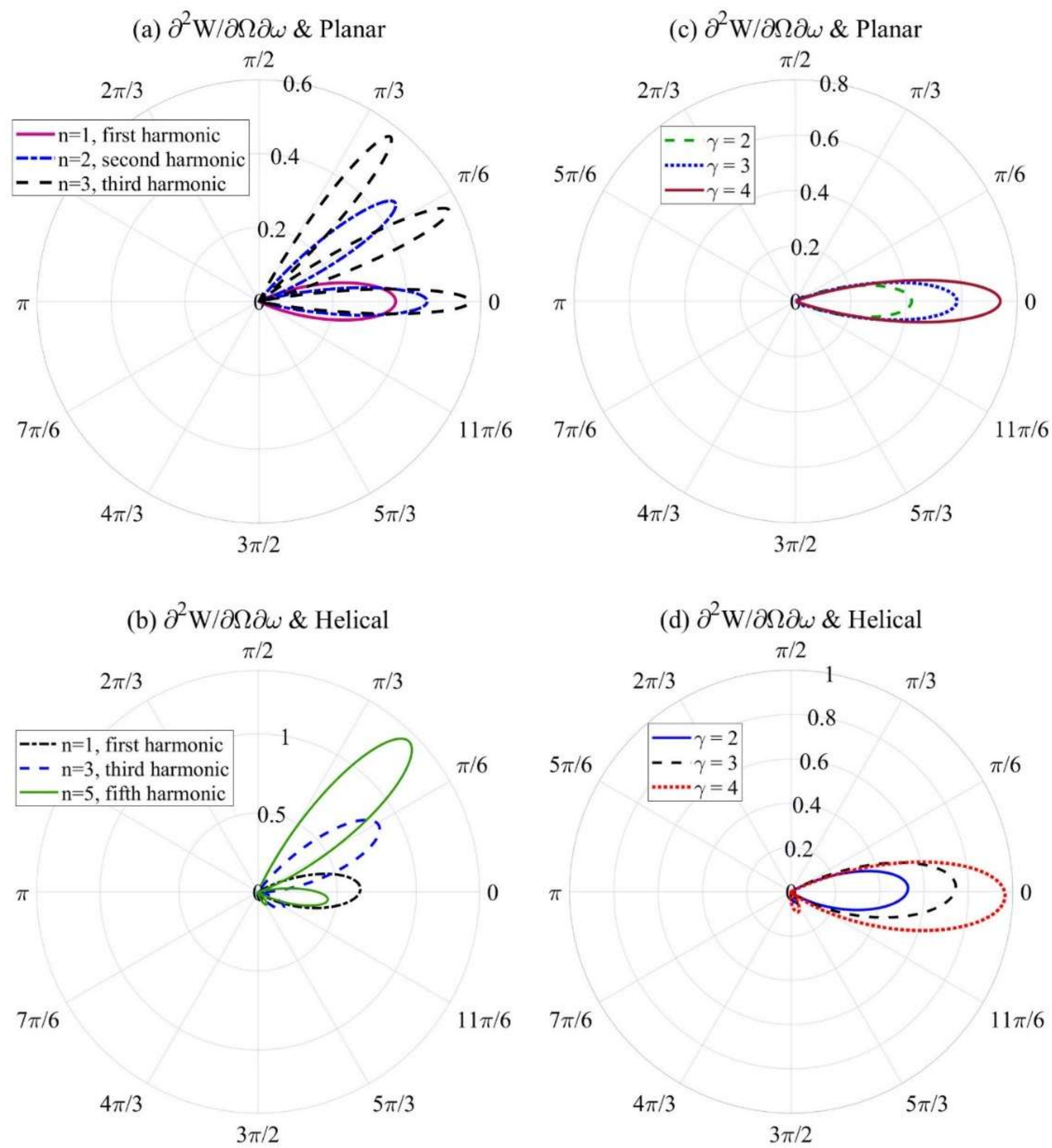

**Fig. 3.** Plots of FEL radiation patterns for various values of the Lorentz factor ($\gamma$) and different harmonic orders in planar and helical undulators.

Figure 4 presents the Fourier-transformed electric field spectrum of the FEL at the fundamental harmonic for different values of the normalized undulator parameter $a_u$. The figure shows a clear, monotonic correlation between increasing undulator strength and both the frequency upshift and intensity enhancement of the emitted radiation. The observed spectral variations arise from fundamental changes in the interaction between the electron beam and the magnetic field within the undulator. As the magnetic field strength increases (corresponding to larger $a_u$), electrons undergo stronger transverse oscillations along their trajectories. Simultaneously, the increased field amplitude effectively lengthens the electron path, shifting the resonant frequency of the emitted radiation. This enhanced coupling also introduces nonlinear effects in the emission process: as the oscillation amplitude grows, electron trajectories deviate from an ideal sinusoidal form. These nonlinearities slightly distort the spectrum, producing side oscillations and a broader envelope around the principal peak at higher undulator strengths. Nevertheless, the spectral peaks remain sharp and well defined, indicating that the radiation retains a high degree of temporal and spatial coherence.

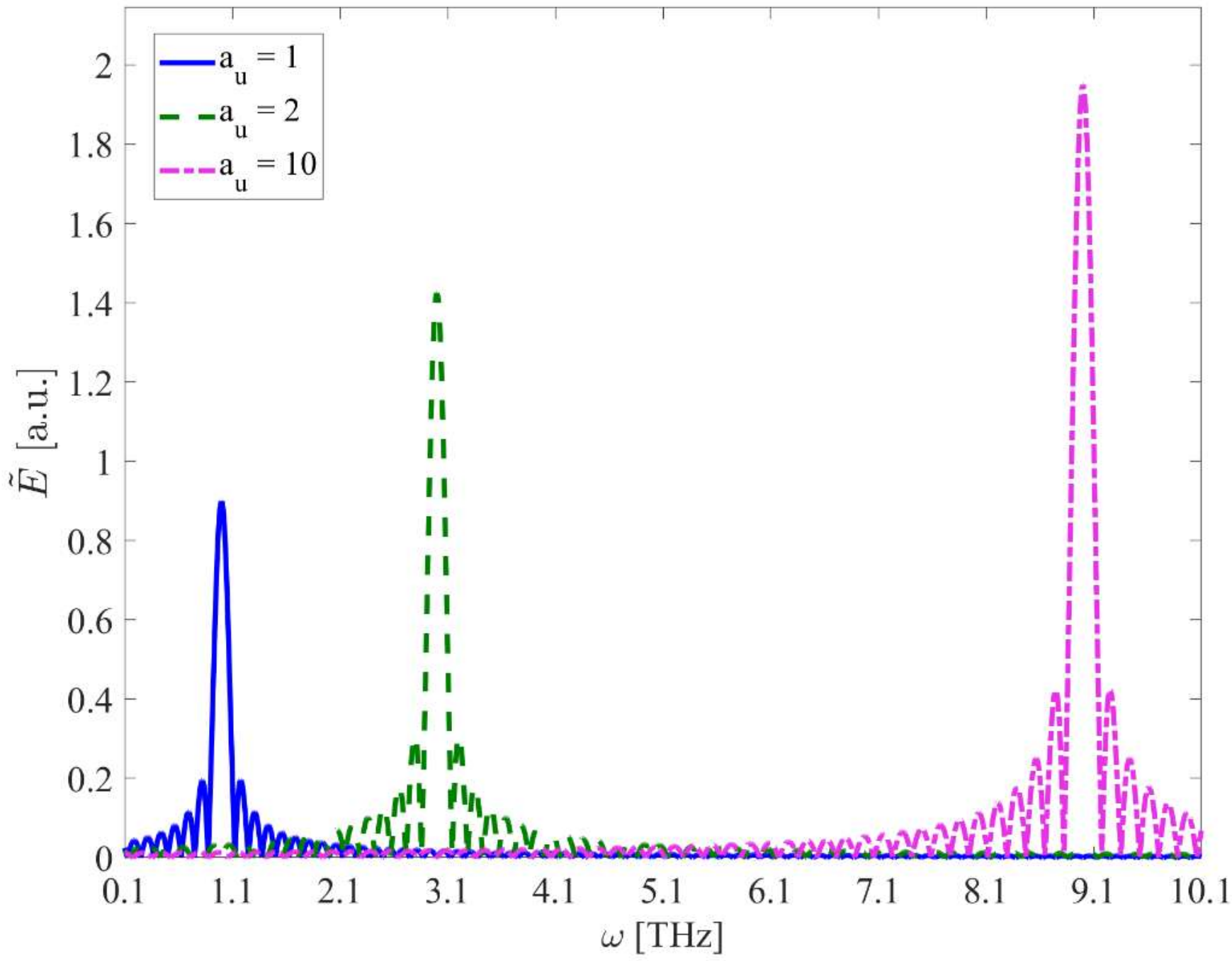

**Fig. 4.** Fourier transform of the FEL electric field for the first harmonic and various values of $a_u$.

To analyze the collective behavior of electron ensembles, Figure 5 shows the dependence of the bunching factor $|b|$ on the beam length $L_z$, normalized by the resonant wavelength $\lambda_r$, for four distinct longitudinal charge distributions: Gaussian, Lorentzian, bi-Gaussian, and pre-bunched. The bunching factor quantifies the phase coherence of electrons within one radiation wavelength and is a key parameter in determining the efficiency of coherent radiation mechanisms such as FELs or THz sources based on modulated beams. The Gaussian distribution exhibits a smooth exponential decay of $|b|$ as beam length increases, consistent with the rapid loss of spatial coherence once the bunch length becomes comparable to or exceeds the resonant wavelength $\lambda_r$. The Lorentzian profile decays more slowly at small $L_z/\lambda_r$ due to its longer tails, which preserve residual coherence slightly better at intermediate lengths, but ultimately follows a similar monotonic suppression trend. The bi-Gaussian distribution, modelled as two symmetrically shifted Gaussians, produces oscillations and nulls in $|b|$ caused by destructive interference between the two peaks. The zero crossings indicate complete cancellation of coherent microbunching due to phase mismatch, a feature that could be exploited for spectral shaping or controlled harmonic suppression in advanced beam manipulation schemes. The pre-bunched distribution reveals the richest harmonic structure, with quasi-periodic oscillations in $|b|$ persisting even at large $L_z/\lambda_r$. These arise from the sinc-modulated envelope of a finite-length train of bunched electrons. The periodicity reflects the Fourier-limited coherence of a well-defined spatial modulation, and the multiple maxima in pre-bunched plot correspond to constructive interference -- a hallmark of pre-modulated beams and a key feature in broadband THz or harmonic generation setups. In addition, for a perfectly bunched beam, $|b|$ approaches 1, but deviates due to finite bunch length and phase slippage. The undulator parameter $a_u$ affects $|b|$ by enhancing transverse motion, leading to stronger bunching at higher $a_u$ via improved resonance.

A comparative analysis of the bunching factor magnitude $|b|$ as a function of the normalized longitudinal bunch length $L_z/\lambda_r$, for various electron beam profiles and undulator parameters $a_u$ is provided in Figure 6. The bunching factor, a key metric in assessing microbunching quality in FEL systems and THz radiation sources, shows strong sensitivity to both the spatial charge distribution and the undulator field strength. For the Gaussian profile (top-left), the bunching factor decays smoothly and monotonically with increasing bunch length due to the exponential suppression of phase coherence. Larger $a_u$ values enhance the initial bunching at small $L_z$, reflecting stronger modulation and effective compression. However, for $L_z/\lambda_r \approx 0.6$, all curves converge toward near-zero bunching, emphasizing the intrinsic limitation of Gaussian beams for longer bunches. The Lorentzian profile (top-right), with its long temporal tails, exhibits a slower decay in $|b|$, maintaining moderate bunching up to $L_z/\lambda_r \approx 0.8$. This suggests that beams with heavy-tailed distributions can sustain microbunching over longer extents, potentially benefiting broadband coherent emission. While higher $a_u$ still enhances field-induced coherence, the decay remains smoother than in the Gaussian case due to the Lorentzian's slower spatial fall-off. For the bi-Gaussian profile (bottom-left), representing a beam with internal modulation or two sub-pulses, $|b|$ displays non-monotonic behaviour. A pronounced dip near $L_z/\lambda_r \approx 0.5$ reflects destructive interference between sub-pulses, followed by a secondary rise from partial constructive rephasing. Although increasing $a_u$ boosts the initial bunching, it cannot fully suppress the interference-induced minimum, underscoring the complexity of using modulated beams for coherent THz generation. In the pre-bunched case (bottom-right), the behavior is markedly oscillatory, with periodic bunching revivals, arising from intrinsic harmonic structure of the pre-bunched beam (modelled via sinc modulation). Increasing $a_u$ not only amplifies these revivals but also sharpens their contrast, reflecting more efficient harmonic bunching. Such dynamics are advantageous for applications targeting narrowband THz or FEL emission, where resonant enhancement of higher harmonics is needed.

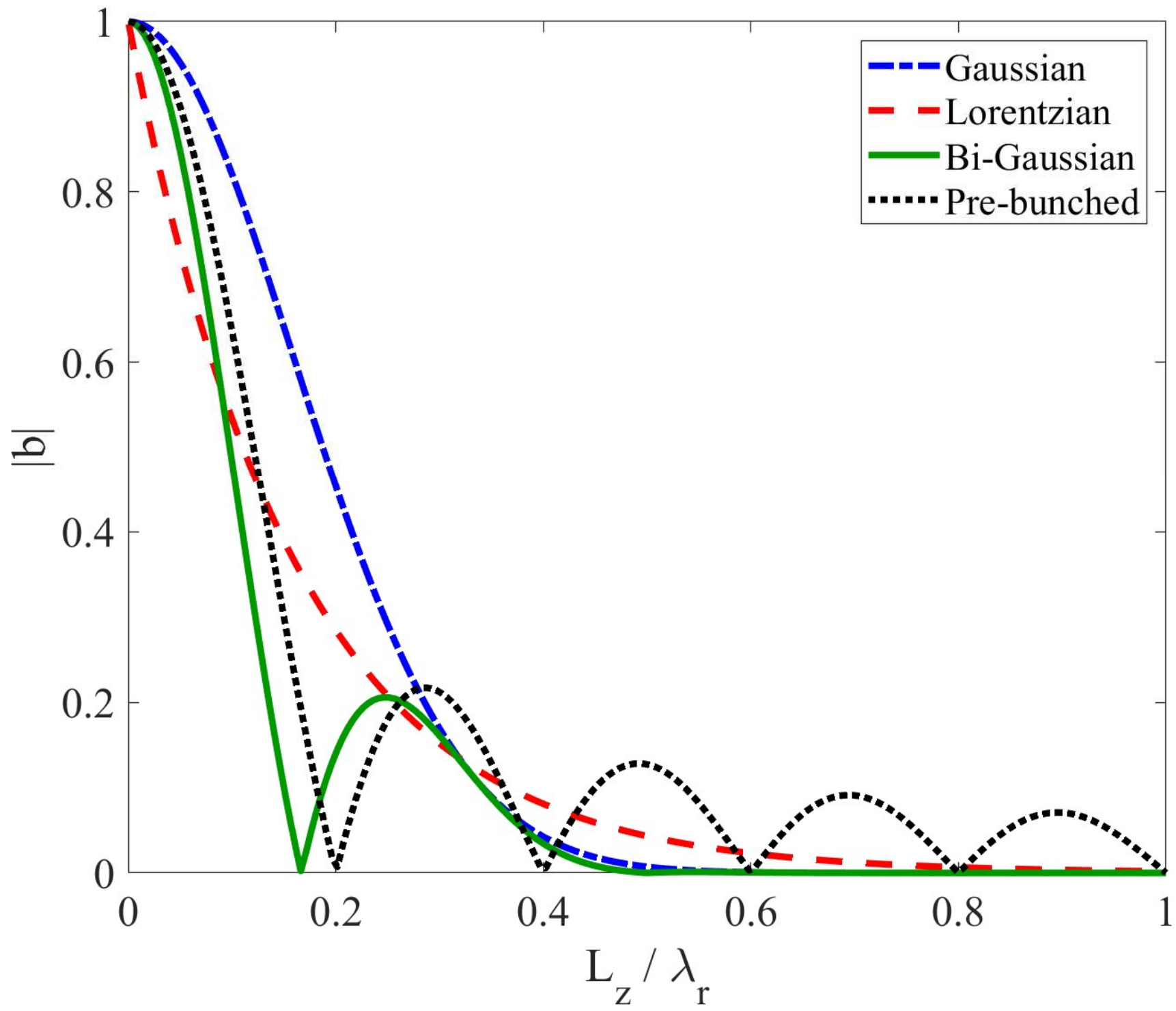

**Fig. 5.** The normalized bunching factor $|b|$ as a function of the normalized longitudinal beam length $L_z/\lambda_r$ for four different electron beam distributions.

Plasma effects are modeled as space-charge–induced dispersion within the electron beam, with densities $n_e = 10^{15} - 10^{17} cm^{-3}$ , which are typical for FEL beams and correspond to normalized values of $\omega_p/\omega_r = 0.01 - 0.2$ in this study. This effective plasma exists longitudinally along the beam as a result of the high charge density, and its magnitude is varied to represent realistic operating conditions in compact THz FELs. Figure 7 explores how the normalized plasma frequency $\omega_p/\omega_r$, which characterizes the collective plasma response of the medium, influences the evolution of the bunching factor |b| as a function of normalized bunch length $L_z/\lambda_r$ for various initial longitudinal beam profiles. Plasma frequency effects are introduced as exponential damping terms reflecting plasma-induced dispersion and space-charge-induced decoherence. For the Gaussian profile (top-left), the bunching factor decays smoothly and exponentially with increasing $L_z$, with higher plasma frequencies inducing faster suppression. At a low ratio of $\omega_p/\omega_r = 0.01$, the beam retains relatively strong bunching over a broad range, but as the ratio rises to 0.2, the decay accelerates sharply. This underscores the vulnerability of Gaussian beams to collective plasma fields, where even modest space-charge effects can significantly degrade phase coherence. In the Lorentzian profile (top-right), the decay is similar but slightly slower. The Lorentzian's long temporal tails help preserve partial bunching at larger $L_z$, though higher $\omega_p/\omega_r$ values still impose strong damping. This suggests that Lorentzian beams offer modest resilience against plasma-induced decoherence, likely due to their extended interaction envelope. For the bi-Gaussian profile (bottom-left), representing a beam with internal structure and phase interference, plasma effects produce a more intricate response. A local minimum in |b| near $L_z/\lambda_r \approx 0.4$, reflects destructive interference between sub-bunches. As $\omega_p/\omega_r$ increases, the overall contrast diminishes, and the interference pattern becomes blurred, illustrating how plasma fields erode internal coherence even in structured beams. The pre-bunched configuration (bottom-right) displays periodic bunching revivals arising from its intrinsic harmonic modulation. However, this coherence is highly sensitive to plasma frequency: even a moderate value of $\omega_p/\omega_r = 0.05$ substantially reduces the bunching peaks, and at $\omega_p/\omega_r = 0.2$, the modulation is nearly suppressed. The damping of higher harmonics by the plasma response imposes a serious constraint on pre-bunched beam performance in high-density environments.

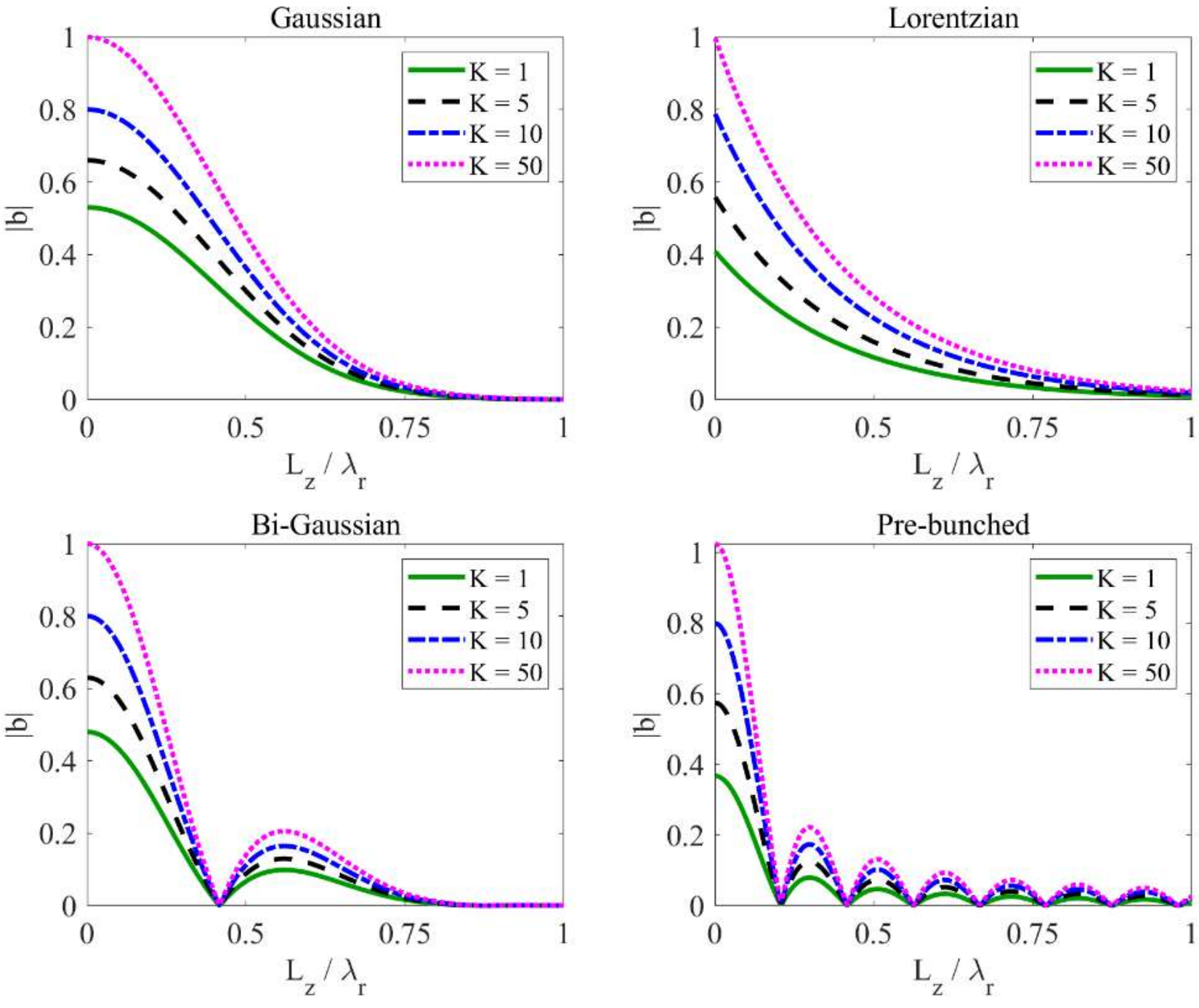

**Fig. 6.** Influence of bunching behaviour on the magnitude of the normalized bunching factor $|b|$ as a function of the normalized bunch length $L_z/\lambda_r$ for four different beam distribution functions and undulator parameters $a_u$.

The spatial structure and beam energy jointly govern the preservation of longitudinal coherence, and thus the potential for high-efficiency coherent radiation. In Figure 8, each subplot presents curves for a range of relativistic Lorentz factors, directly linking the radiation wavelength to the beam energy. Since the resonant wavelength scales inversely with $\gamma^2$, increasing the beam energy effectively shortens the scale over which longitudinal decoherence occurs. In the Gaussian and Lorentzian panels, the bunching factor decreases monotonically with $L_z/\lambda_r$, with the decay rate steepening as $\gamma$ increases. This trend reflects the decreasing radiation wavelength at higher energies; for the same physical bunch length, the normalized parameter $L_z/\lambda_r$ becomes larger, amplifying phase decorrelation among electrons. Lorentzian profiles exhibit a slightly slower initial decay than Gaussian ones, consistent with their longer statistical tails, which sustain phase correlation over a broader range. The shape-preserving decay in both cases indicates that for smooth, unimodal profiles, the dominant mechanism is the exponential dephasing. The Bi-Gaussian distribution behaves qualitatively differently. Two spatially separated charge peaks introduce interference fringes in the bunching factor, particularly evident at higher $\gamma$, where phase mismatch is more severe. Oscillatory minima, corresponding to destructive interference, shift toward shorter normalized lengths as $\lambda_r$ decreases. This illustrates the strong phase sensitivity of bimodal distributions and suggests possible strategies for tailoring emission through controlled longitudinal modulation. The pre-bunched case displays even more pronounced oscillations, arising from the Fourier-limited spectral response of a periodic train of electron bunches. As $\gamma$ increases, the oscillation frequency in $|b|$ rises due to greater phase accumulation per unit length. Notably, even for relatively long bunches ($L_z/\lambda_r > 0.5$), substantial bunching persists at lower $\gamma$, demonstrating the robustness of pre-bunching schemes at moderate energies. In contrast, high $\gamma$ beams require significantly shorter modulation lengths to maintain microbunching, owing to faster phase slippage. Furthermore, the resonance wavelength $\lambda_r$ is related to beam energy by $\lambda_r = \lambda_u(1 + a_u^2)/(2\gamma^2)$. GENESIS simulations confirm the expected quadratic dependence of coherent energy on charge ($\propto N_b^2|b|^2$), as evidenced by parabolic scaling observed in output power versus charge scans, in good agreement with theory.

The spectral characteristics of radiation from different electron beam distributions show pronounced sensitivity to both spatial profile and relativistic energy, as illustrated in Figure 9. For Gaussian beams, a sharp and symmetric spectral peak emerges, but it broadens and weakens with increasing Lorentz factor due to the reduced coherence length at shorter resonant wavelengths. This effect directly limits spectral purity at high beam energies, a critical constraint for applications such as narrowband THz generation or precision pump-probe experiments. Lorentzian beams produce broader spectra with extended tails, maintaining higher intensity at off-resonant frequencies. This indicates greater tolerance to phase slippage and dephasing, making them advantageous for broadband radiation schemes that require coherence across wide spectral ranges. The Bi-Gaussian distribution exhibits interference-induced spectral splitting, with twin peaks that converge and weaken as $\gamma$ increases. This behavior arises from the diminishing relative phase coherence between the spatially separated density peaks, underscoring the strong energy dependence of interference-driven radiation. In contrast, pre-bunched beams generate a rich harmonic spectrum, with regularly spaced peaks whose spacing and sharpness both increase with energy.

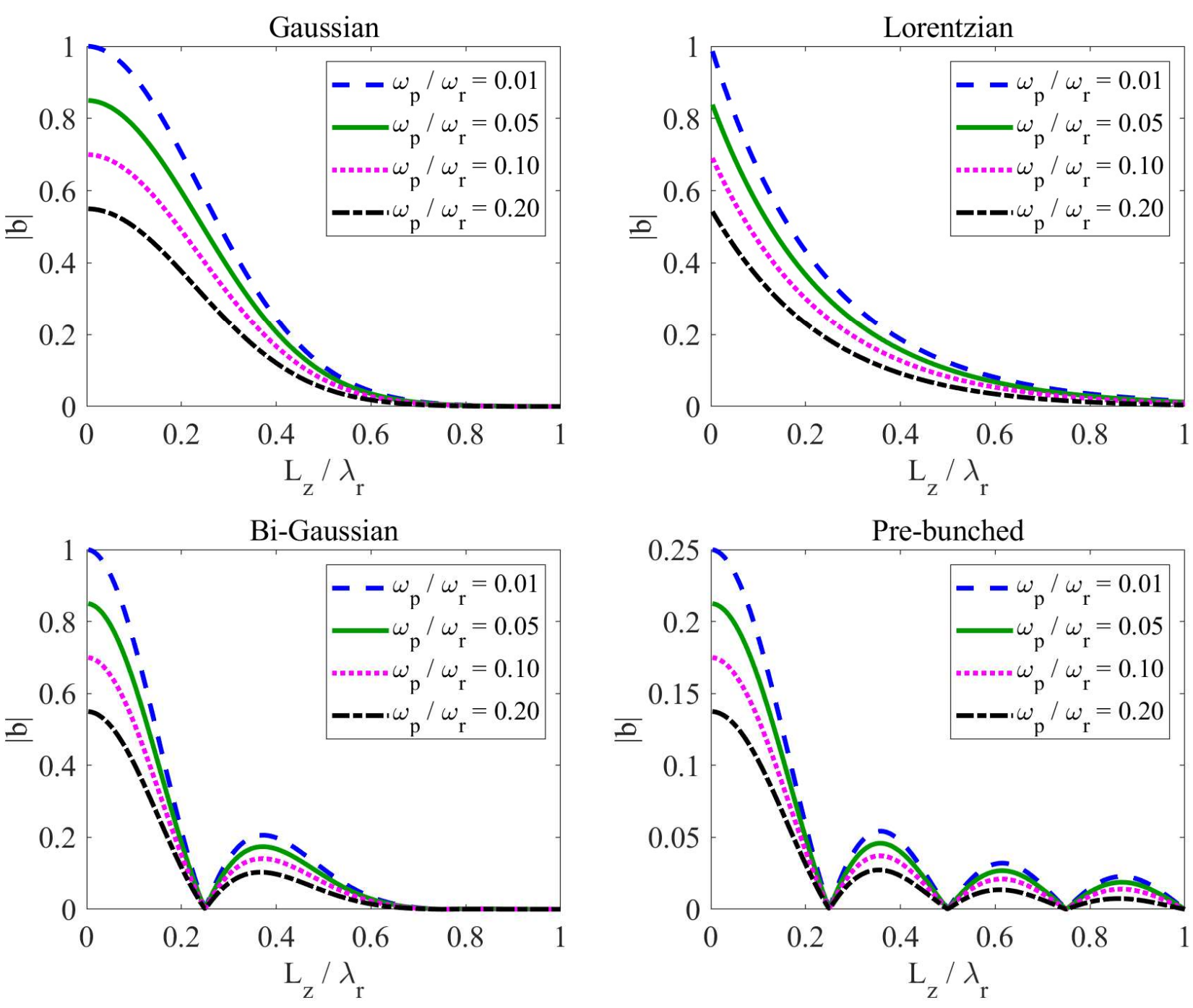

**Fig. 7.** Variation of the normalized bunching factor $|b|$ with normalized bunch length $L_z/\lambda_r$ for four different electron beam distributions and plasma frequency ratios ($\omega_p/\omega_r$).

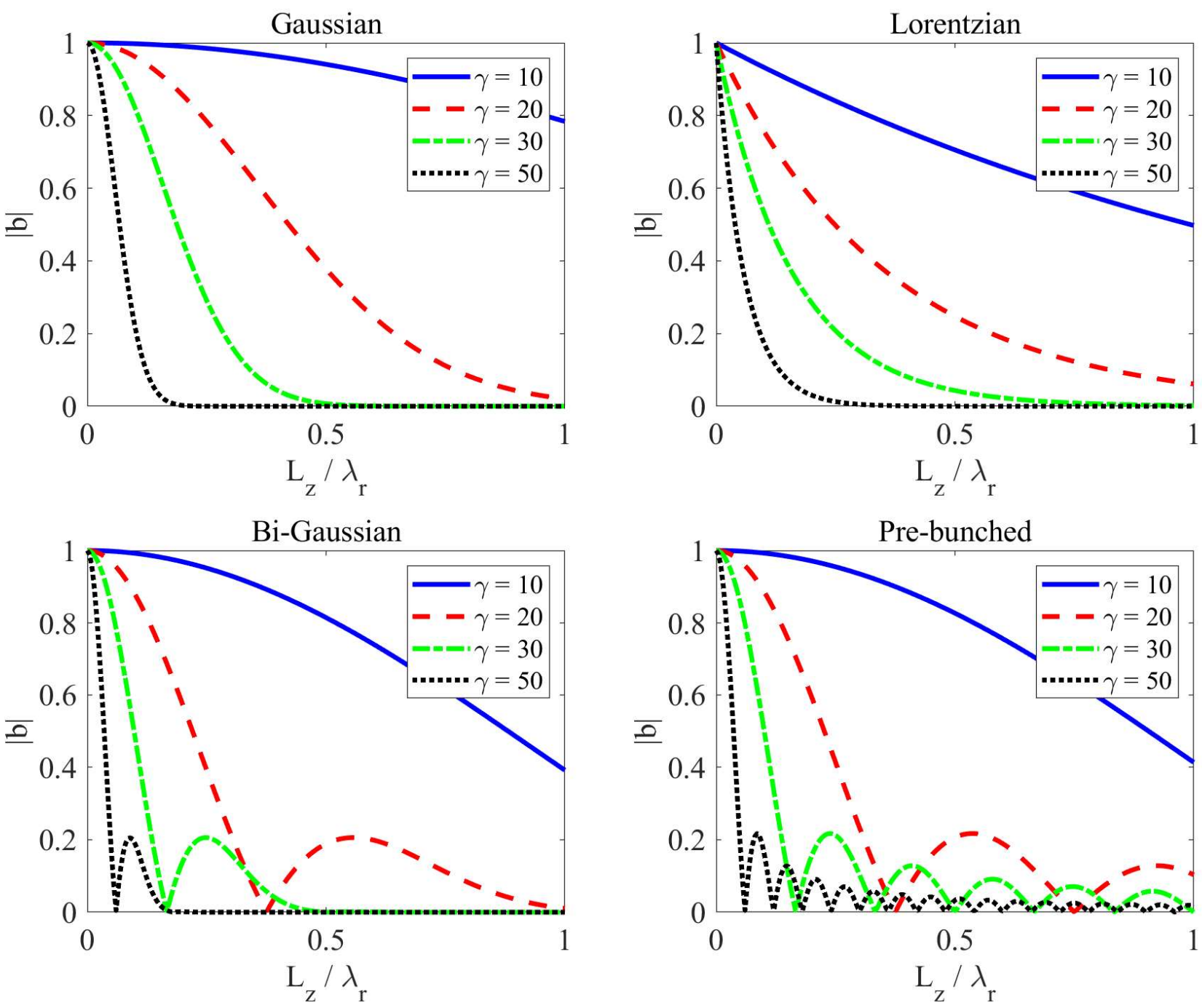

**Fig. 8.** Effect of relativistic factors on the magnitude of the normalized bunching factor |b| as a function of the normalized bunch length $L$ for four different beam distribution functions.

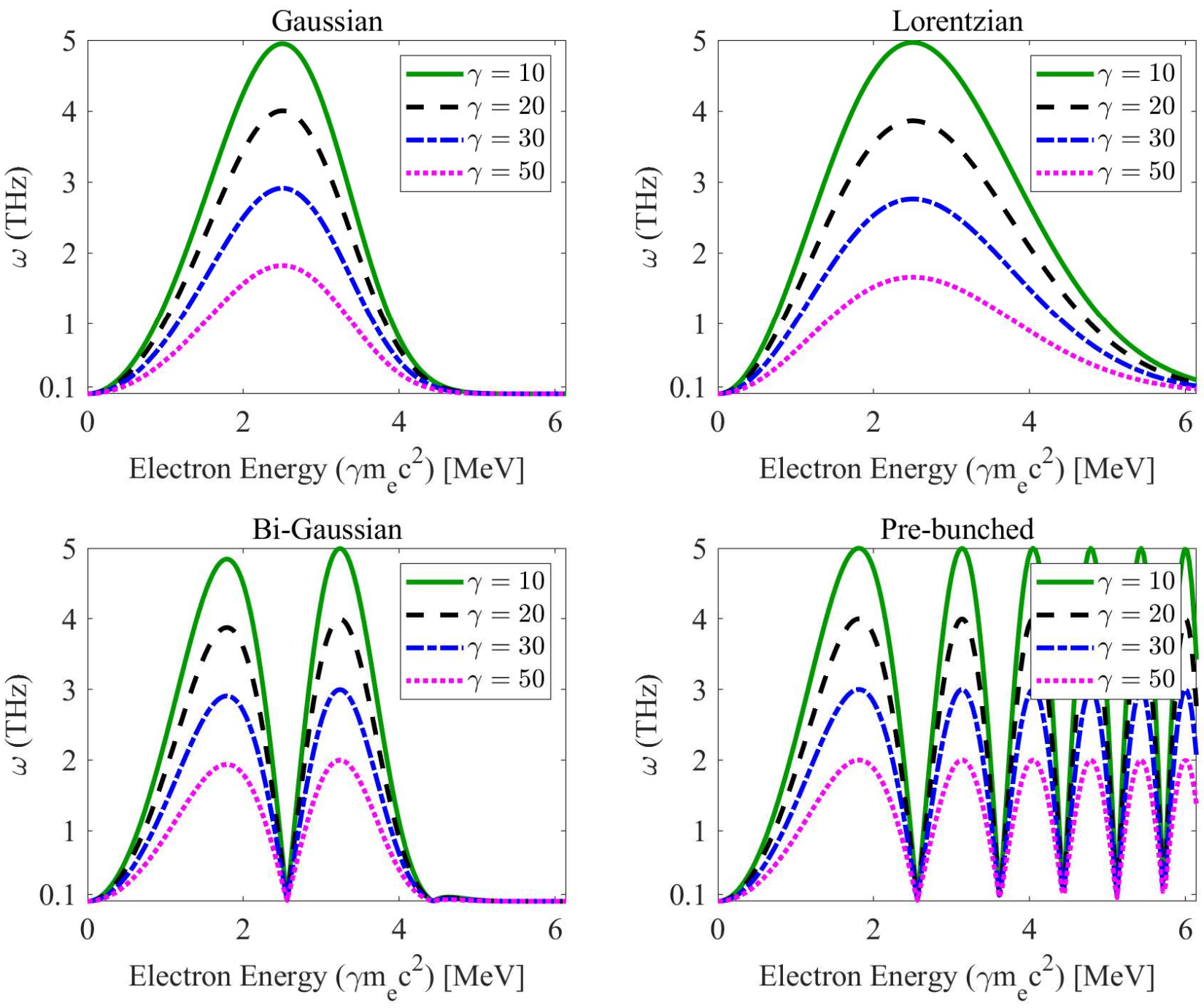

**Fig. 9.** Spectral intensity of electron beam radiation in the THz range as a function of relativistic energy (MeV) for different beam distribution functions across a range of Lorentz factors.

The dependence of the normalized radiation intensity as a function of undulator parameter $a_u$ for four distinct longitudinal charge distributions and relativistic Lorentz factors is presented in Figure 10. In undulator-based systems, radiation intensity is strongly influenced by both the electron beam's longitudinal profile and the undulator parameter $a_u$, as these jointly determine phase coherence and harmonic content of the emitted field. For Gaussian beams, intensity rises rapidly with increasing $a_u$ before reaching saturation at moderate values. This saturation plateau is slightly reduced at higher Lorentz factors due to diminished effective bunching, indicating a loss of phase coherence at shorter resonant wavelengths. In the Lorentzian case, radiation intensity grows more steeply at low $a_u$ and transitions gradually into a flatter high-$a_u$ regime. The extended distribution tails contribute to broader response, enabling efficient emission even strong undulator fields, a robustness that may benefit high-flux operation or broadband photon generation. The Bi-Gaussian profile introduces a more complex intensity pattern, with oscillations superimposed on the overall growth. These modulations arise from phase interference between the two Gaussian components and become smoother at higher $\gamma$, where shorter wavelengths cause the interference fringes to narrow. This pronounced sensitivity to beam microstructure could be exploited for tailoring radiation spectra or optimizing mode purity in advanced light sources. Pre-bunched beams display a distinctly different behavior: a staircase-like intensity curve composed of discrete peaks and plateaus. As $\gamma$ increases, the spacing between these features decreases, reflecting tighter phase locking and enhanced harmonic selectivity.

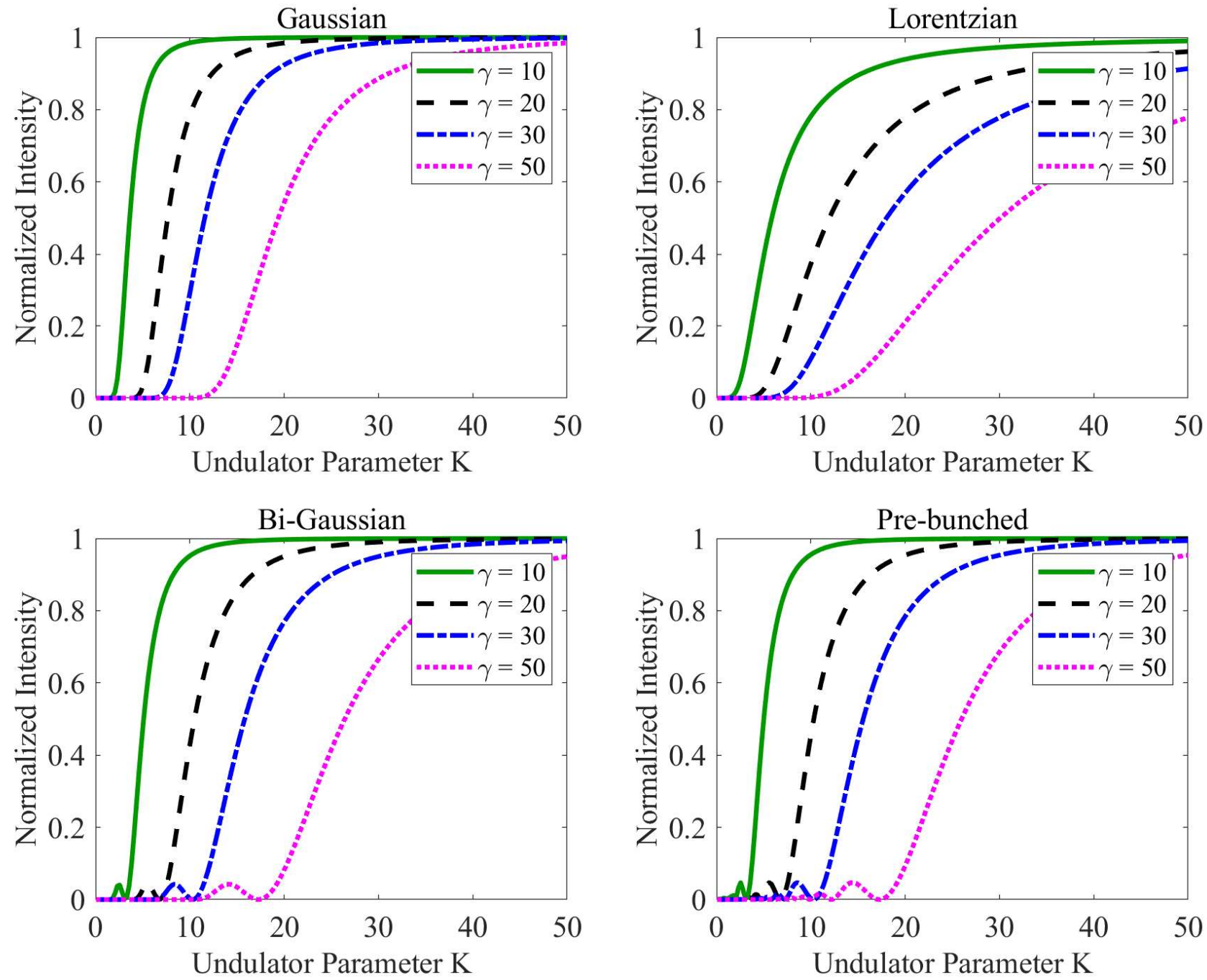

**Fig. 10.** Influence of Lorentz factors on the normalized radiation intensity as a function of the undulator parameter for different beam distributions.

The dependence of radiation intensity on normalized bunch length $L_z/\lambda_r$ is evaluated in Figure 11 for four representative electron beam distributions under varying modulation indices $N$. The analysis reveals how spatial coherence and harmonic content evolve with both bunch geometry and the number of modulation periods, providing a quantitative basis for optimizing coherent emission in advanced accelerator-based light sources. For Gaussian beams, intensity rises sharply to a maximum near $L_z/\lambda_r$, before decaying smoothly as coherence diminishes. Higher modulation indices yield narrower, more intense peaks, consistent with enhanced microbunching, while lower $N$ values result in broader, weaker profiles, reflecting reduced spectral sharpness at shallow modulation depths. This trend directly impacts spectral resolution and is critical for high-precision applications. Lorentzian distributions display a similar initial rise but with longer tails, particularly at high $N$, due to their broader spatial structure. The sustained intensity at larger $L_z/\lambda_r$ indicates a greater tolerance to bunch lengthening, which can be advantageous in dynamic or noisy beam conditions. The bi-Gaussian profile introduces oscillatory modulations from interference between two spatial peaks. These oscillations are most pronounced at high $N$, where strong phase correlations exist, and gradually vanish at low $N$, indicating the loss of well-defined periodicity. Such structured emission could be exploited for tailored light shaping or multi-band generation. Pre-bunched beams display periodic intensity maxima, resulting from constructive interference at discrete harmonic conditions. As $N$ decreases, these peaks broaden and shift, and the spectrum transitions from sharp harmonic features to smoother profiles.

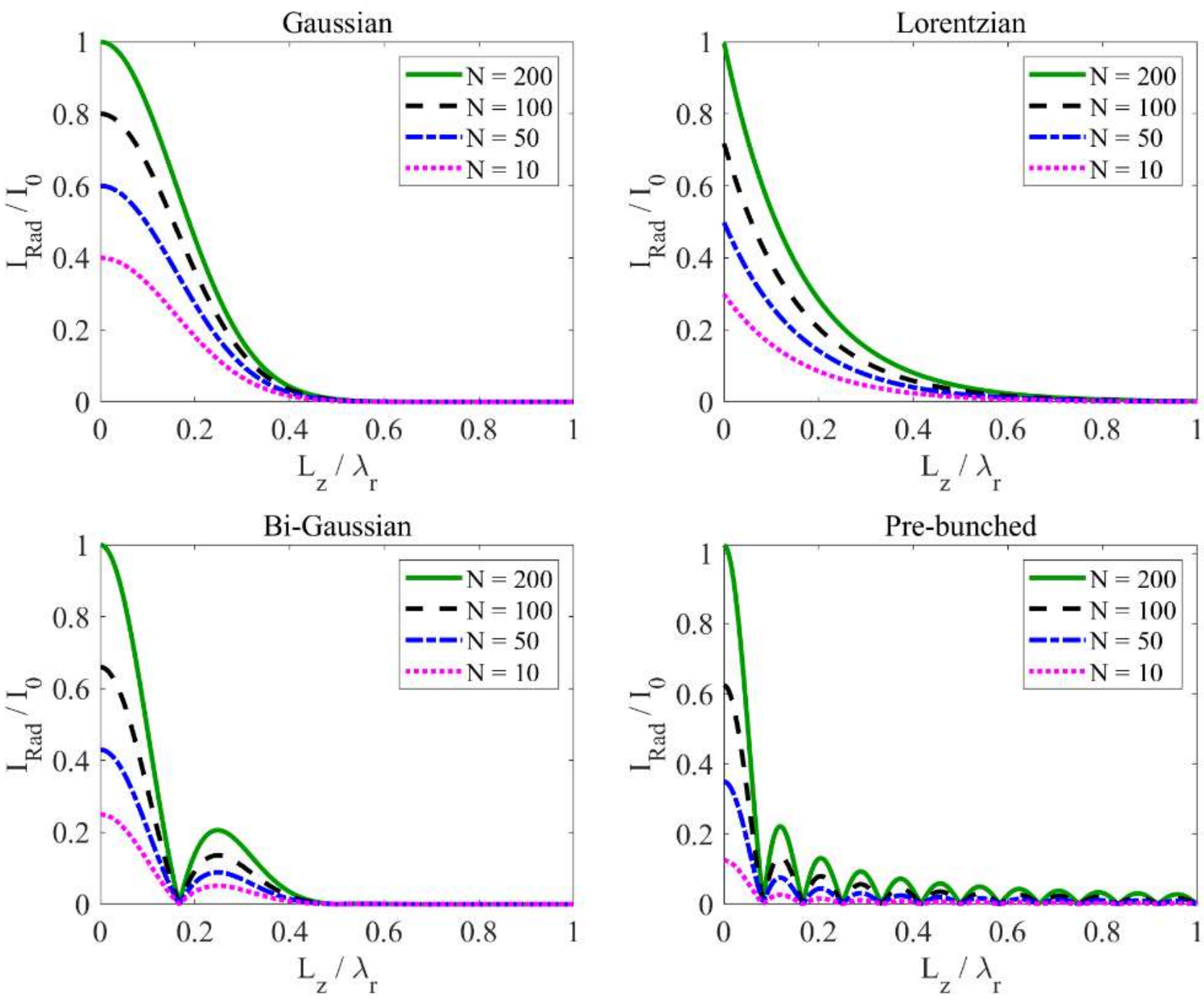

**Fig. 11.** Variation of the normalized radiation intensity as a function of the normalized bunch length ($L_z/\lambda_r$) for various beam distributions over a range of modulation indices.

Coherence degradation from energy spread is a critical limiting factor in high-brilliance light sources. Figure 12 quantifies the relationship between the bunching factor $|b|$ and normalized radiation intensity as function of relative energy spread. At low energy spreads, $|b|$ remains close to unity, indicating strong phase synchronization across the electron ensemble and optimal conditions for coherent emission. As $\Delta\gamma/\gamma$ increases, $|b|$ rapidly declines below 0.1, marking the onset of an incoherent regime. The bunching factor, computed with the time-dependent GENESIS 1.3 code [27, 28] and shown in this Figure. GENESIS simulation markers, with associated uncertainties, confirm the theoretical trend while revealing stochastic deviations from beam imperfections and numerical sampling—features essential for evaluating real-world beamline performance. The normalized radiation intensity follows a similar decline, consistent with its quadratic dependence on |b|. The steep drop imposes a practical upper limit on energy spread for efficient harmonic generation, a threshold of particular importance for time-resolved THz spectroscopy, attosecond pulse production, and THz-based structural probes, where spectral purity depends on maintaining strong microbunching. The close agreement between theory and GENESIS results underscores the model’s predictive power and the importance of energy spread control in FEL design. Mitigation strategies include energy chirp correction and beam phase-space linearization to delay coherence loss. The analysis also provides a diagnostic benchmark for characterizing FEL beam quality and guiding simulation–experiment feedback in the optimization next-generation light sources.

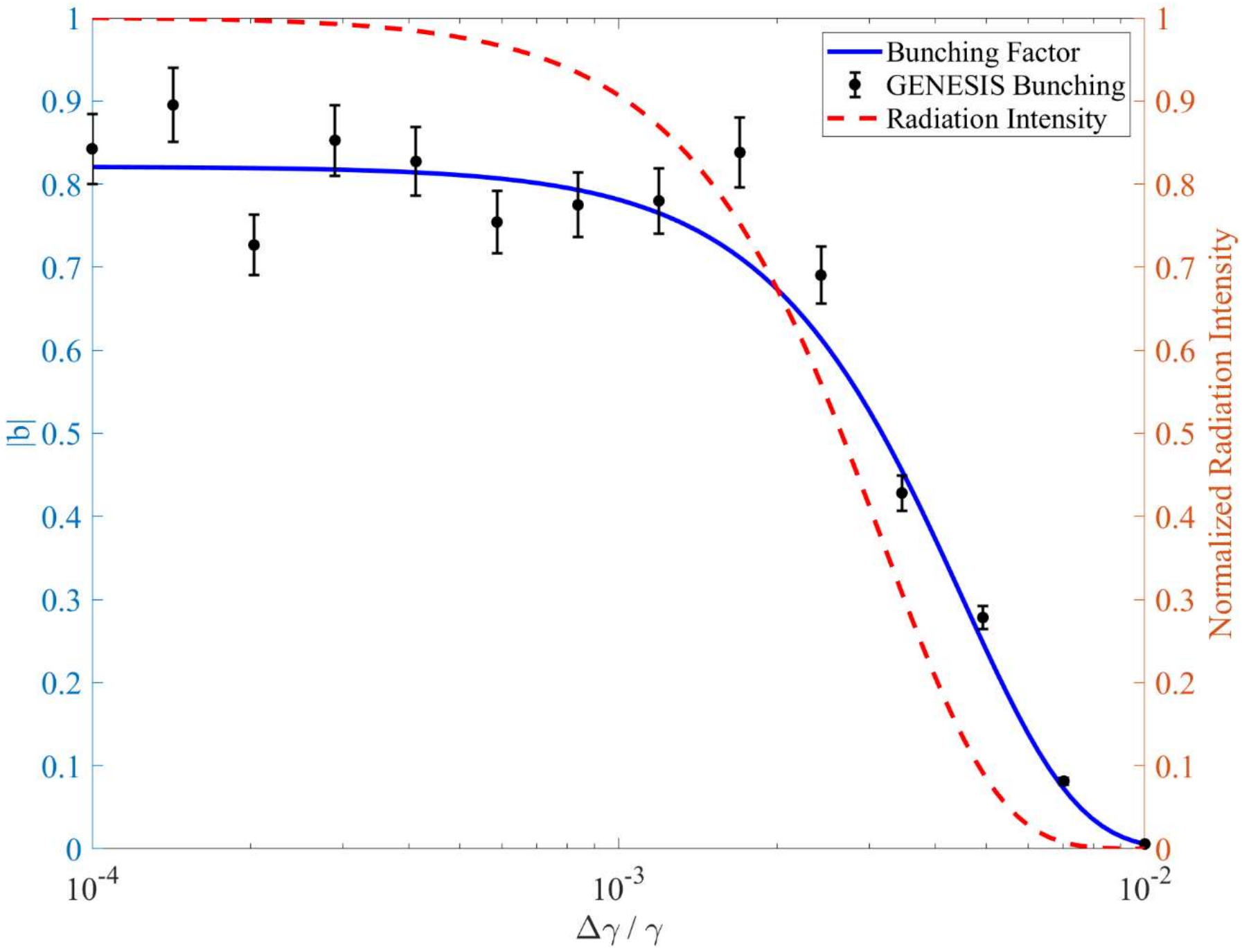

**Fig. 12.** Analysis of the bunching factor and normalized radiation intensity as functions of the relative energy spread, including theoretical predictions and GENESIS simulation results.

## 4. Conclusions

This work presents a combined theoretical and numerical study of terahertz (THz) radiation mechanisms in free-electron laser (FEL) systems, integrating Liénard–Wiechert field analysis with GENESIS 1.3 simulations. Using generalized Bessel function expansions, the radiation electric field and the double differential spectral-angular distribution ($\partial^2 W_{THz}/\partial\omega\partial\Omega$) were derived, establishing a rigorous framework to assess how electron beam and undulator parameters shape THz emission. The results reveal that both planar and helical undulators can produce coherent THz radiation but with distinct angular and polarization characteristics. Planar undulators exhibit stronger sidelobes and broader angular spreads, especially at higher harmonics, whereas helical undulators generate more collimated emission owing to their continuous magnetic symmetry. Regardless of geometry, the radiation tends to be strongly forward-directed, with polarization determined by the combined properties of the beam and magnetic field configuration rather than the undulator shape alone. Transitioning into the wiggler regime at high magnetic parameters broadens the angular profile and introduces lateral lobes, highlighting the need for field optimization to preserve spectral purity. Beam profile strongly influences coherence and harmonic generation. Pre-bunched beams achieve the highest harmonic conversion efficiency but are highly sensitive to energy spread and plasma-induced dispersion, requiring advanced beam conditioning techniques. Lorentzian beams offer greater resilience to decoherence over wide energy spreads, while bi-Gaussian profiles produce controllable interference modulations useful for spectral shaping. GENESIS simulations agree closely with analytical predictions, validating the framework and quantifying the interplay between the bunching factor, energy spread, and radiation intensity. Slippage effects are inherently included in GENESIS through the velocity mismatch between the electrons and the radiation; however, no additional velocity-reducing components (such as waveguides) were introduced, ensuring that the dynamics reflect only those of the undulator. These insights provide concrete design guidelines for next-generation high-brilliance THz sources. In particular, they emphasize controlling energy spread, optimizing undulator configuration, and tailoring beam profiles to the target application. The results are directly relevant to attosecond science, nonlinear optics, ultrafast spectroscopy, and other fields where spectral purity and coherence are paramount.

**Appendix A:**

Evaluation of integral in Eq. 6:

$$e^{i\omega\left(t-\frac{\hat{n}.\vec{r}}{c}\right)} = e^{i\omega Ht} e^{iA}$$

$$H = 1 - \beta cos\theta, \quad \beta = 1 - \frac{\left(1+\frac{a_{u'}^{2}}{2}\right)}{2\gamma^2}$$

$$\mathrm{A}{=}d\sin(\omega_u t) + \sigma\cos\left(\omega_u t\right)$$

$$d = \frac{a_{u'} sin\,\phi sin\theta}{\gamma\omega_u}, \qquad \sigma = \frac{a_{u'}\omega cos\phi cos\theta}{\gamma\omega_u}$$

$$e^{iA} = \sum_n e^{-in\omega_u t}\Delta_n$$

$$\Delta_n = e^{-in\left(\frac{u}{2}\right)} J_n\left(-\sqrt{u^2-d^2}, 0; \exp\left(iu\right)\right)$$

$$u = tg^{-1}(2tg\phi)$$

**Acknowledgments.** This research did not receive any specific grant from funding agencies in the public, commercial, or not-for-profit sectors.

**Disclosures.** The authors declare no conflicts of interest.

**Data availability.** The data that support the findings of this study are available from the corresponding author upon reasonable request.

## References

1. J. F. Zhu, Ch. H. Du, T. J. Huang, L. Y. Bao, Sh. Pan, P. K. Liu, “Free-electron-driven beam-scanning terahertz radiation,” Optics Express, 27, 18, (2019), https://doi.org/10.1364/OE.27.026192.
2. W. Wang, P. K. Lu, A. K. Vinod, D. Turan, J. F. McMillan, H. Liu, M. Yu, D. L. Kwong, M. Jarrahi, Ch. W. Wong, “Coherent terahertz radiation with 2.8-octave tunability through chip-scale photo-mixed micro-resonator optical parametric oscillation,” Nature Communications, 13, 1, (2022), https://doi.org/10.1038/s41467-022-32739-6.
3. H. Ge, Zh. Sun, Y. Jiang, X. Wu, Zh. Jia, G. Cui, Y. Zhang, “Recent Advances in THz Detection of Water,” International Journal of Molecular Sciences, 24, 13, (2023), https://doi.org/10.3390/ijms241310936.
4. J. Jin, H. Xiong, J. Zhou, M. Guang, X. Wu, “Strong-field THz radiation-induced curing of composite resin materials in dentistry,” Biomedical Optics Express, 14, 5, (2023), https://doi.org/10.1364/BOE.484241.
5. J. D. Yuzon, Z. Schultzhaus, Zh. Wang, “Transcriptomic and genomic effects of gamma-radiation exposure on strains of the black yeast Exophiala dermatitidis evolved to display increased ionizing radiation resistance,” Microbiology spectrum, 11, 5, (2023), https://doi.org/10.1128/spectrum.02219-23.
6. T. Nörenberg, G. Á. Pérez, M. Obst, L. Wehmeier, F. Hempel, J. M. Klopf, A. Y. Nikitin, S. C. Kehr, L. M. Eng, P. A. González, Th. V. A. G. de Oliveira, “Germanium Monosulfide as a Natural Platform for Highly Anisotropic THz Polaritons,” ACS nano, 16, 12, (2022), https://doi.org/10.1021/acsnano.2c05376.
7. H. La, A. Brokkelkamp, S. V. D. Lippe, J. T. Hoeve, J. Rojo, S. C. Boj, “Edge-induced excitations in Bi2Te3 from spatially-resolved electron energy-gain spectroscopy,” Ultramicroscopy, 254, 113841, (2023), https://doi.org/10.1016/j.ultramic.2023.113841.
8. T. Tanaka, “Difference frequency generation in free electron lasers,” Optics letters, 43, 18, (2018), https://doi.org/10.1364/OL.43.004485.
9. Ch. Feng, X. Wang, T. Lan, M. Zhang, X. Li, J. Zhang, W. Zhang, L. Feng, X. Liu, H. Deng, B. Liu, D. Wang, Zh. Zhao, “Slippage boosted spectral cleaning in a seeded free-electron laser,”Scientific reports, 9, 1, (2019), https://doi.org/10.1038/s41598-019-43061-5.
10. R. Pompili, D. Alesini, M. P. Anania, S. Arjmand, M. Behtouei, M. Bellaveglia, A. Biagioni, B. Buonomo, F. Cardelli, M. Carpanese, E. Chiadroni, A. Cianchi, G. Costa, A. D. Dotto, M. D. Giorno, F. Dipace, A. Doria, F. Filippi, M. Galletti, L. Giannessi, A. Giribono, P. Iovine, V. Lollo, A. Mostacci, F. Nguyen, M. Opromolla, E. D. Palma, L. Pellegrino, A. Petralia, V. Petrillo, L. Piersanti, G. D. Pirro, S. Romeo, A. R. Rossi, J. Scifo, A. Selce, V. Shpakov, A. Stella, C. Vaccarezza, F. Villa, A. Zigler, M. Ferrario, “Free-electron lasing with compact beam-driven plasma wakefield accelerator,” Nature, 605 (7911), 659-662, (2022), https://doi.org/10.1038/s41586-022-04589-1.
11. M. Galletti, D. Alesini, M. P. Anania, S. Arjmand, M. Behtouei, M. Bellaveglia, A. Biagioni, B. Buonomo, F. Cardelli, M. Carpanese, E. Chiadroni, A. Cianchi, G. Costa, A. D. Dotto, M. D. Giorno, F. Dipace, A. Doria, F. Filippi, G. Franzini, L. Giannessi, A. Giribono, P. Iovine, V. Lollo, A. Mostacci, F. Nguyen, M. Opromolla, L. Pellegrino, A. Petralia, V. Petrillo, L. Piersanti, G. D. Pirro, R. Pompili, S. Romeo, A. R. Rossi, A. Selce, V. Shpakov, A. Stella, C. Vaccarezza, F. Villa, A. Zigler, M. Ferrario, “Stable Operation of a Free-Electron Laser Driven by a Plasma Accelerator,” Physical review letters, 129 (23), 234801, (2022), https://doi.org/10.1103/PhysRevLett.129.234801.
12. W. Liu, Y. Lu, L. Wang and Q. Jia, “A compact terahertz free-electron laser with two gratings driven by two electron-beams,” Physics of Plasmas, 24, 023109, (2017), https://doi.org/10.1063/1.4976122.
13. S. C. Sharma, J. Panwar and R. Sharma, “Modeling of terahertz radiation emission from a free electron laser,” Contrib. Plasma Phys. 2017;57, (2017), https://doi.org/10.1002/ctpp.201600085.
14. G. Q. Liao, H. Liu, G. G. Scott, Y. H. Zhang, B. J. Zhu, Zh. Zhang, Y. T. Li, Ch. Armstrong, E. Zemaityte, Ph. Bradford, D. R. Rusby, D. Neely, P. G. Huggard, P. McKenna, C. M. Brenner, N. C. Woolsey, W. M. Wang, Zh. M. Sheng and J. Zhang, “Towards Terawatt-Scale Spectrally Tunable Terahertz Pulses via Relativistic Laser-Foil Interactions,” Physical Review x, 10, 031062 (2020), https://doi.org/10.1103/PhysRevX.10.031062.
15. K. Floettmann, F. Lemery, M. Dohlus, M. Marx, V. Tsakanov and M. Ivanyan, “Superradiant Cherenkov–wakefield radiation as THz source for FEL facilities,” J. Synchrotron Rad., 28, 18–27, (2021), https://doi.org/10.1107/S1600577520014058.
16. M. Lenz, A. Fisher, A. Ody, Y. Park, and P. Musumeci, “Electro-optic sampling-based characterization of broad-band high efficiency THz-FEL,” Optics Express, 30, 19, (2022), https://doi.org/10.1364/OE.467677.
17. E. Roussel, Ch. Szwaj, C. Evain, B. Steffen, Ch. Gerth, B. Jalali and S. Bielawski, “Phase Diversity Electro-optic Sampling: A new approach to single-shot terahertz waveform recording,” Light Sci Appl 11, 14 (2022), https://doi.org/10.1038/s41377-021-00696-2.
18. A. Fisher, Y. Park, M. Lenz, A. Ody, R. Agustsson, T. Hodgetts, A. Murokh and P. Musumeci, “Single-pass high-efficiency terahertz free-electron laser,” Nature Photonics, 16, (2022), https://doi.org/10.1038/s41566-022-00995-z.
19. V. Petrillo, A. Bacci, I. Drebot, M. Opromolla, A. R. Rossi, M. R. Conti, M. Ruijter, S. Samsam and L. Serafini, “Synchronised Terahertz Radiation and Soft X-rays Produced in a FEL Oscillator,” Appl. Sci., 12, 8341, (2022), https://doi.org/10.3390/app12168341.
20. K. Zhukovsky, I. Fedorov, N. Gubina, “Theoretical analysis of the influence of electron beam parameters on the harmonic powers in free electron lasers,” Optics & Laser Technology, 159, 108972, (2023), https://doi.org/10.1016/j.optlastec.2022.108972.

21. L. Feigin, A. Gover, A. Friedman, A. Weinberg, D. Azar and A. Nause, "High-Power Terahertz Free Electron Laser via Tapering-Enhanced Superradiance," Electronics, 13, 1171, (2024), https://doi.org/10.3390/electronics13071171.
22. A. Fisher, M. Lenz, A. Ody, Y. Yang, Ch. Pennington, J. Maxson, T. Hodgetts, R. Agustsson, A. Murokh, P. Musumeci, "Towards higher frequencies in a compact prebunched waveguide THz-FEL," Nature Communications, 15:7582, (2024), https://doi.org/10.1038/s41467-024-51892-8.
23. J. D. Jackson, "Classical Electrodynamics," New York, Wiley, ISBN: 9780471309321, (1999).
24. D. Babusci, G. Dattoli, S. Licciardi, E. Sabia, "Mathematical methods for physicists," World Scientific Publishing, (2020).
25. Y. Tian, J. Liu, Y. Bai, Sh. Zhou, H. Sun, W. Liu, J. Zhao, R. Li, Zh. Xu, "Femtosecond-laser-driven wire-guided helical undulator for intense terahertz radiation," Nature Photonics, 11, 4, (2017), https://doi.org/10.1038/nphoton.2017.16.
26. A. A. Molavi Choobini and F. M. Aghamir, "Effects of multi-color femtosecond laser beams and external electric field on transition-Cherenkov THz radiation," Phys. Plasmas 29, 103106, (2022), https://doi.org/10.1063/5.0087840.
27. S. Reiche, "GENESIS 1.3: a fully 3D time-dependent FEL simulation code," Nucl. Instrum. Methods Phys. Res., Sect. A, 429, 243, (1999), https://doi.org/10.1016/S0168-9002(99)00114-X.
28. Genesis, https://github.com/svenreiche/Genesis-1.3-Version4.